\documentclass{article}

\usepackage{plain}

\date{}
\begin{document}

\title{The impact of the \textit{storytelling fallacy} on real data examples \\ in methodological research}

\author[1,2]{\normalsize Maximilian M. Mandl\footnote{\textbf{Correspondence:} Maximilian M. Mandl, Institute for Medical Information Processing, Biometry, and Epidemiology, Faculty of Medicine, Ludwig-Maximilians-Universität München, Marchioninistr. 15, 81377 München, Germany. mmandl@ibe.med.uni-muenchen.de}}
\author[1]{\normalsize Frank Weber}
\author[3]{\normalsize Tobias Woehrle}
\author[1,2]{\normalsize Anne-Laure Boulesteix}

\affil[1]{\small Institute for Medical Information Processing, Biometry and Epidemiology, Faculty of Medicine, Ludwig-Maximilians-Universität München, Germany}
\affil[2]{\small Munich Center for Machine Learning, Germany.}
\affil[3]{\small Department of Anesthesiology, LMU University Hospital, Ludwig-Maximilians-Universität München, Germany}

\renewcommand\Authands{ and }
  \maketitle

\begin{abstract}
The term \lq\lq researcher degrees of freedom'' (RDF), which was introduced in metascientific literature in the context of the replication crisis in science, refers to the extent of flexibility a scientist has in making decisions related to data analysis. These choices occur at all stages of the data analysis process, e.g., data preprocessing and modelling. In combination with selective reporting, RDF may lead to over-optimistic statements and an increased rate of false positive findings. Even though the concept has been mainly discussed in fields of application of statistics such as epidemiology or psychology, similar problems affect methodological statistical research. Researchers who develop and evaluate statistical methods are left with a multitude of decisions when designing their comparison studies. This leaves room for an over-optimistic representation of the performance of their preferred method(s) and false positive findings. In this context, the present paper defines and explores a particular RDF that has not been previously identified and discussed.  When interpreting the results of real data examples that are most often part of methodological evaluations, authors typically tell a domain-specific \lq\lq story'' that best supports their argumentation in favor of their preferred method. However, there are often plenty of other plausible stories that would support different conclusions.
We define the \lq\lq storytelling fallacy'' as the selective use of anecdotal domain-specific knowledge to support the superiority of specific methods in real data examples.
While such examples fed by domain knowledge play a vital role in methodological research, if deployed inappropriately they can also harm the validity of conclusions on the investigated methods. The goal of our work is to create awareness for this issue, fuel discussions on the role of real data in generating evidence in methodological research and warn readers of methodological literature against naive interpretations of real data examples. We illustrate this newly introduced RDF through two examples related to pleiotropy detection in Mendelian Randomisation and a prediction model to detect SARS-CoV-2 infections, respectively.

\end{abstract}


\newpage
\section{Introduction}
\label{section_1}

The term \lq\lq researcher degrees of freedom'' (RDF) \cite{simmons2011false}, which was introduced in metascientific literature in the context of the replication crisis in science \cite{baker20161, gelman2014statistical, loken2017measurement}, refers to the extent of flexibility a scientist has in making decisions related to data analysis. These choices occur at all stages of the data analysis process, e.g., data preprocessing and modelling \cite{hoffmann2021multiplicity}. In combination with selective reporting, RDF may lead to over-optimistic statements and an increased rate of false positive findings \cite{hoffmann2021multiplicity,ioannidis2005most}. Even though the concept has been so far mainly discussed in fields of application of statistics such as epidemiology or psychology, similar problems affect methodological statistical research. Here, we define methodological statistical research as research dedicated to the development and evaluation of statistical methods, for example statistical tests or regression modelling---as opposed to research addressing epidemiological questions using these methods.

Researchers developing and evaluating statistical methods are left with a multitude of decisions when designing their comparison studies.
Jelizarow et al. \cite{jelizarow2010over} demonstrate how a new class prediction method that is in fact worse than existing methods can artificially seem superior through selective reporting. More precisely, they intentionally focus on the datasets, data preprocessing settings, and method variants that maximize the performance of the new method while downplaying others. This selective reporting approach leads to an overoptimistic and misleading representation of their preferred method. Such bias induced by the RDF typically affects the evaluations presented as part of papers introducing new methods, including those based on statistical simulation studies \cite{ullmann2023over,pawel2024pitfalls}. 

It should be emphasized that we do not insinuate that scientists deliberately engage in scientific misconduct. In fact, selective reporting often happens subconsciously without malicious intention, as a result of self-deception \cite{nuzzo2015fooling}. The reasons are manifold. One key factor that has been widely acknowledged in health and social sciences is publication bias that encourages the reporting of positive findings. It is likely that such a bias also affects methodological literature  \cite{boulesteix2015publication}---in the sense that researchers are required to propose methods that outperform existing methods to have their research published.

In this context, the present paper defines and explores a particular RDF that has not been previously identified and discussed.  When interpreting the results in real data examples that are most often part of methodological evaluations, authors typically tell a domain-specific \lq\lq story'' that supports their argumentation in favor of their preferred method. However, there are often plenty of other plausible stories that would support different conclusions.
We define the \lq\lq storytelling fallacy'' as the selective use of anecdotal domain-specific knowledge to support the superiority of specific methods in real data examples.
While applications fed by domain knowledge play a vital role in methodological research, if deployed inappropriately they can also harm the validity of conclusions on the investigated methods and lead to the publication of non-replicable methodological results. 

The goal of our paper is to create awareness for this issue, fuel discussions on the role of real data in generating evidence in methodological research and warn readers of methodological literature against naive interpretations of real data examples. 
After introducing the concept of the \lq\lq storytelling fallacy'' in more detail in Section~\ref{section_2},  we illustrate it through two examples inspired from our own research. In Section~\ref{section_3}, we consider methods for pleiotropy detection in Mendelian Randomisation (MR) for causal inference. The different methods yield different sets of results. For each of these sets of results, a plausible biological interpretation (\lq\lq story'') can be elaborated to strengthen the case of the corresponding method. In Section~\ref{section_4}, we consider a new diagnostic method for SARS-CoV-2 infections based on machine learning models. Different models identify different predictor variables as relevant. Again, for each set of results, a plausible biological \lq\lq story'' based on prior knowledge on these predictor variables can be created to support the use of the corresponding model. The paper finally discusses these findings and formulates recommendations for authors and readers of methodological comparison studies with respect to real data examples (Section~\ref{section_5}).

\section{The \lq\lq storytelling fallacy''}
\label{section_2}
\subsection{Definition}
Building on Jelizarow et al. \cite{jelizarow2010over}, Nießl et al. \cite{niessl2022over} systematically investigate the impact of RDF on the results of a showcase study comparing the accuracy of various survival prediction methods. Typical RDF included in their study are the choices of performance measures, datasets, summarization of results over datasets, and handling of method failures. Depending on these choices, different survival prediction methods can be identified as best performing. Assuming that authors often engage (consciously or not) in selective reporting, this may largely explain why papers introducing new methods are generally over-optimistic with respect to their performances, even if they use essentially objective criteria \cite{niessl2024explaining,boulesteix2013plea,buchka2021optimistic}.

In this context, a natural reaction would be to give more importance to real data applications and to the plausibility of the results obtained therein from a domain-knowledge perspective. If---according to expert judgement derived from domain knowledge---the results obtained with method A are much more plausible and meaningful than those obtained with method B, it can be seen as an argument in favor of the superiority of method A. 

Such interpretations are common in methodological articles presenting new methods. For example, let us consider the case of a new model selection approach for multivariable regression modelling (called method A and compared to a standard approach called method B). In a real data application, the inventors of method A who want to present it as superior to method B may scrutinize the variables it selects and argue that they are more meaningful than those selected by method B. They may for example argue that method A selects a variable that was identified as important predictor of the target variable in previous literature, and that method B fails to select it. This would be an argument in favor of the superiority of method A.

The goal of the present paper is to outline that RDF also affect this type of comparisons in a broad sense, and that the interpretation of real data applications is consequently not immune against bias. This is because, for a given real data application, there are usually numerous possible sensible domain-specific interpretations. Focusing on the one that makes method A appear better than method B in some sense (while ignoring those that make method B look better) can be seen as a form of selective reporting. This happens in a particularly subtle way, because in practice the authors do not actively select their interpretation out of a collection of ready-to-use interpretations: instead, they elaborate their interpretation based on the results of methods A and B---but may have elaborated another one if their preferences had been different.

With this in mind, we define the \lq\lq storytelling fallacy'' as the use of anecdotal domain-specific knowledge in order to support the superiority of the preferred method in real data examples. 

\subsection{Related literature}
The \lq\lq storytelling fallacy'' is related to the question of the reliability and objectivity of domain-specific expert knowldege, which has been widely discussed in different fields of research. We give a brief overview of this literature in connection with the \lq\lq storytelling fallacy'' in the rest of this section.

Experts are generally assumed to be immune to blind spots and generally impartial \cite{dror2020cognitive} and, as stated by O'Hagan \cite{o2019expert}, \lq\lq \textit{[e]xpert opinion and judgment enter into the practice of statistical inference and decision-making in numerous ways. Indeed, there is essentially no aspect of scientific investigation in which judgment is not required.}'' Regarding subjectivity, they further argue that \lq\lq \textit{[j]udgment is necessarily subjective, but should be made as carefully, as objectively, and as scientifically as possible.}''
Kaptchuk \cite{kaptchuk2003effect} points out that the interpretation of data is inevitably subjective and is not preserved from bias---referred to as interpretative bias. Furthermore, the well-known confirmation bias \cite{oswald2004confirmation} refers to the fact that researchers may evaluate evidence in a way that supports their own prior beliefs. 

In statistics, the notion of expert judgement is often considered in the debate opposing frequentist to Bayesian statistics, which often ends in discussions on the subjectivity and objectivity of decisions \cite{gelman2017beyond,brownstein2018perspective} such as the choice of the prior distributions in Bayesian statistics. 
The \lq\lq storytelling fallacy'' in methodological literature is related to the subjectivity of expert judgement, but in a different manner. The judgement of an expert who elaborates an interpretation to strengthen the argument in favor of a newly introduced method is subjective---in the sense that another expert may have another judgement. However, the core issue is the RDF combined with selective reporting, i.e. the multiplicity of potential post-hoc stories and the fact that the one that best fits the authors' hopes is chosen, leading to a biased evaluation of the methods.

With a different perspective, Boutron and Ravaud \cite{boutron2018misrepresentation} address the problem of misrepresentation of research in biomedical literature and discuss the notion of  \lq\lq spin''. The spin is defined as a type of reporting that does not accurately represent the nature and scope of findings and may influence readers' perceptions of the results. There are different shades of \lq\lq spins'', such as the misreporting of methods and the misreporting of results. 
Similarly, methodological researchers may evaluate their results in a biased way by focusing on specific performance measures (i.e. methods), datasets (i.e. results) or, as outlined in this paper, specific biological theories supporting the veracity of results of real data analyses and thus the superiority of the method that produced them. \\

\section{Use Case I: Detection of pleiotropic SNPs in Mendelian Randomisation}
\label{section_3}

\begin{figure*}
\centerline{\includegraphics[scale=0.25]{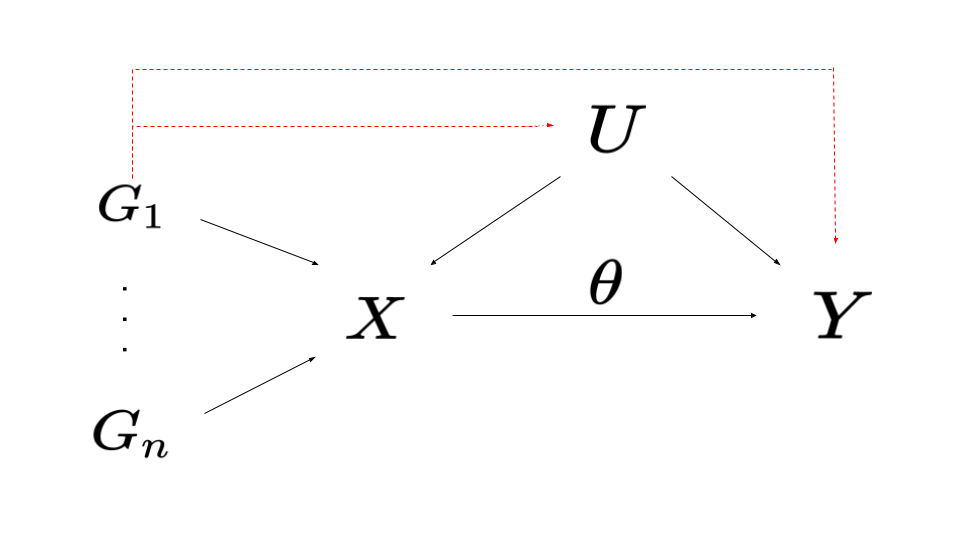}}
\caption{Causal directed acyclic graph (DAG) for univariable MR, based on Mandl et al. \cite{mandl2025outlierdetectionmendelianrandomisation}. Genetic variants are denoted as $G_i$ for $i \in 1,...,n$, confounders as $U$, and the causal effect of the risk factor $X$ on the outcome $Y$ as $\theta$. Red dashed lines represent the effect of the instrumental variable(s) $G_i$ on the outcome through paths other than the risk factor---e.g., caused by pleiotropy. \label{figure_1}}
\end{figure*}

Mendelian Randomisation (MR) employs genetic variants as instrumental variables to deduce the causal effects of exposures on an outcome. A crucial assumption in MR is that these genetic variants, used as instrumental variables, are independent of the outcome, given the risk factor and any unobserved confounders.\cite{burgess2013mendelian}
More precisely, the goal of MR is to examine the causal effect $\theta$ of a risk factor $X$ on an outcome $Y$ using genetic variants $G_i$ for $i \in 1,...,n$ as instrumental variables (IV), see Figure~\ref{figure_1}. Pleiotropy is defined as the effect of any genetic variant $G_i$ (IV) on the outcome $Y$ through any other path than the risk factor $X$ included in the MR model---see the red dashed lines in Figure~\ref{figure_1}. Different approaches to detect pleiotropic SNPs exist in the literature \cite{verbanck2018detection, sanderson2019examination, bowden2015egger}. They are mainly based on adjusted versions of Cochran's Q.

In this section, we revisit parts of the results of our recent study comparing methods for the detection of pleiotropic SNPs \cite{mandl2025outlierdetectionmendelianrandomisation} to illustrate the \lq\lq storytelling fallacy''. Beyond extensive statistical simulations, this comparison study includes a real data application assessing the effect of circulating vitamin D levels (exposure: $X$) on multiple sclerosis (MS, outcome: $Y$) through MR; see the original paper for details on data and methods \cite{mandl2025outlierdetectionmendelianrandomisation}. Table~\ref{table_1} shows the results of two different pleiotropy detection methods: Method A (Standard) and method B (GC-Q). Method A detects one pleiotropic SNP, namely \textit{rs4944958} while method B does not detect any pleiotropic SNPs.\footnote{Details on code and data for the analysis can be found on GitHub: \url{https://github.com/mmax-code/MR_outliers}}

\begin{center}
\begin{table}[hbt!]
\centering
\begin{adjustbox}{width=8cm,center}
\begin{tabular}{lc}
\hline
\noalign{\vskip 5pt}
Method & pleiotropic SNP  \\[1ex] 
\hline\hline
\noalign{\vskip 5pt}
Method A (Standard)       & rs4944958 \\
Method B (GC-Q)           &  ---  \\
\bottomrule
\end{tabular}
\end{adjustbox}
\caption{Real data analysis from \cite{mandl2025outlierdetectionmendelianrandomisation}: pleiotropic SNPs for the analysis of vitamin D on MS in the univariable MR analysis.}
\label{table_1}
\end{table}
\end{center}

\begin{center}
\begin{table}[hbt!]
\centering
\begin{tabular}{p{7.5cm}  p{7.5cm}}
\hline 
\centering \makecell{\textbf{Story 1} \\ Method A (Standard)} & \makecell{\textbf{Story 2} \\ Method B (GC-Q)}   \\[1ex] 
\hline\hline 
\noalign{\vskip 5pt}
SNPs associated with vitamin D are known to belong to three tiers, the direct vitamin D pathway, U/V absorption, and the cholesterol metabolism \cite{manousaki2020genome}. The pleiotropic SNP identified by method~A, \textit{rs4944958}, is an intron of the \textit{NADSYN1} gene which affects a precursor of cholesterol and is thus involved in cholesterol metabolism---which can be seen as confounder for MS \cite{murali2022}. Therefore, it may indicate horizontal pleiotropy. &
The SNP identified by method~A, \textit{rs4944958}, is considered a perfect proxy for \textit{rs12785878} \cite{mokry2015vitamin}---a SNP that has been linked directly to vitamin D serum concentrations by several studies. Therefore, \textit{rs4944958} has been explicitly included in similar studies, see, e.g, \cite{mokry2015vitamin}. Even though \textit{rs4944958} is involved in the cholesterol metabolism, the effect through this causal path on MS is not finally clarified \cite{lorincz2022}.  Thus, it makes sense to include \textit{rs4944958} in the MR analysis and consider it a false-positive finding of method A. \\
\bottomrule 
\end{tabular}
\caption{Two different domain-specific \lq\lq stories'' in favor of two different methods for the MR use case.}
\label{table_2}
\end{table}
\end{center}

Table~\ref{table_2} shows two different interpretations (\lq\lq stories'') that could be used to argue in favor of the superiority of the methods A and B, respectively, using prior biological domain knowledge. This clearly illustrates that the interpretation of real data results is uncertain or, in other words, affected by RDF, and should be considered with caution.

\section{Use Case II: Breath sample analysis by semi-conductor based E-Nose technology}
\label{section_4}

In order to distinguish SARS-CoV-2 infected from non-infected patients, an experimental analytical approach (E-Nose technology) can be applied, where volatile organic molecules, also termed gasotransmitters, contained in human exhaled air are enriched and analysed by 10 different metal oxide semiconductor sensors \cite{woehrle2024point}. Subsequently, these variations produce signal patterns that can be analysed using machine learning (ML) methods to detect SARS-CoV-2 infections. See the original paper for more details \cite{woehrle2024point}.\footnote{Details on the code of the original analysis can be found on GitHub: \url{https://github.com/mmax-code/enose}}

Two prediction models with SARS-Cov-2 status as dependent variable and different sensor covariates are obtained using a Random Forest (RF: method A) and a Support Vector Machine (SVM: method B) \cite{hastie01statisticallearning}, respectively. Note that for each sensor different covariates were extracted due to the complexity of the raw data. For the purpose of model interpretability, the covariates are then assessed using permutation importance measures. Table~\ref{table_3} shows the three top-ranked sensor-covariates according to the RF- and SVM-based variable importances, respectively. Even though sensor 9  is in the top three ranks for both methods, the other sensors differ.

\begin{table}[h!]
\centering
\begin{adjustbox}{width=9cm,center}
\begin{tabular}{c c c}
\hline
Ranking    & Method A     &  Method B  \\
& (RF) & (SVM) \\
\hline
\noalign{\vskip 5pt}
1.    &  Sensor 9, covariate\textsubscript{9-3}     & Sensor 2, covariate\textsubscript{2-7}     \\
2.    &  Sensor 9, covariate\textsubscript{9-8}     & Sensor 10, covariate\textsubscript{10-1}    \\
3.    &  Sensor 9, covariate\textsubscript{9-2}      & Sensor 9,  covariate\textsubscript{9-1}   \\
\hline
\end{tabular}
\end{adjustbox}
\caption{Top three ranked sensor-covariates according to the permutation importance measure for methods A and B. Note that due to the complexity of the raw sensor data different covariates were extracted for each sensor. The first index of the covariate corresponds to the sensor, while the second stands for the specific covariate extracted from this sensor.}
\label{table_3}
\end{table}

\begin{center}
\begin{table}[hbt!]
\centering
\begin{tabular}{p{7.5cm}  p{7.5cm}}
\hline 
\centering \makecell{\textbf{Story 1} \\ Method A (RF)} & \makecell{\textbf{Story 2} \\  Method B (SVM)}   \\[1ex] 
\hline\hline 
\noalign{\vskip 5pt}
 Method A only outputs sensor 9 in the first three ranks, which detects aromatic and sulphor organic compounds \cite{woehrle2024point}. It is a highly sensitive broad range sensor, that cannot identify a single marker molecule alone. Sensor 9 is especially sensitive for sulphor compounds, as well as the manufacturer’s specification of $H_2S$ as their calibration gas. Like hydrogen and methane, $H_2S$ represents a well-described gasotransmitter in lung disease \cite{jiang2024gasotransmitter}, and endogenous $H_2S$ production in humans may be increased to counteract viral infection and inflammation \cite{citi2020anti}. These arguments are in line with the ranking output by method A. & 

Method B identified sensor 2, which is a very sensitive broad range sensor, and sensor 10 which is selective for methane. Sensors identified by method B hint at methane and hydrogen as potential biomarkers of an underlying SARS-CoV-2 infection. Methane producing microbes can generate methane ($CH_4$) from carbon dioxide ($CO_2$) and hydrogen ($H$ or $H_2$), often encountered in anoxic environments. Thus, elevated levels of hydrogen may facilitate increased production of methane, and both hydrogen and methane have been described as so-called gasotransmitters---small gas molecules that are endogenously generated, have well-defined functions, and play a role in respiratory diseases \cite{jiang2024gasotransmitter}. These arguments are in line with the ranking output by method B. \\
\bottomrule 
\end{tabular}
\caption{Two different domain-specific \lq\lq stories'' in favor of two different methods for the E-Nose use case.}
\label{table_4}
\end{table}
\end{center}

As outlined in Table~\ref{table_4}, it is possible to interpret the results in such a way that method A appears to yield more plausible results. But it is also possible to make method B appear superior.
As in Section~\ref{section_3} we can thus again use domain knowledge to favor one or the other method.

These examples show that despite identifying different sensors, both variable importance measures may have revealed valuable sensors for the detection of pathological processes during infections caused by SARS-CoV-2. From a medical point of view, both approaches yield highly interesting results, and especially in the medical setting with high inter-individual variations, several strategies should be considered and potentially combined, rather than focusing on a single analytical strategy.
The results of the two methods are different, but not necessarily incoherent. The fallacy lies solely in the {\it selective} reporting of arguments in favor of one of the methods.

\section{Discussion}
\label{section_5}

This manuscript discusses a new type of RDF that has not been described before. It was already well-known that methodological comparison studies can be biased in favor of the authors' preferred method(s) through the selective reporting of, e.g., specific datasets or performance measures \cite{niessl2022over}. We argue that the selective reporting of \lq\lq stories'' based on domain knowledge that make the results of the preferred method(s) seem more reliable than those of other methods also contributes to present a biased picture of the methods' qualities.

This raises the question on the role of real data applications in methodological research in general and as piece of evidence on the behavior of methods in particular. We do not claim that real data applications are meaningless. In fact, they have a pivotal role in the various \lq\lq phases of methodological research'' \cite{heinze2024phases}. In an early phase study presenting a new method idea, real data applications may be used to demonstrate that the method can be applied to real data and which type of results it yields. In a late phase study whose goal is to generate reliable evidence of the behavior of a method in various contexts, real data applications may be used to discuss special cases in which the method shows a particular behavior. In this context, it may also make sense to interpret the results based on multiple real data examples along with the results of simulation studies---where the ground truth is known, as done in a late phase study addressing regularized regression methods in combination with graphical Gaussian models \cite{kramer2009regularized}.

If the biological plausibility of the results is considered a crucial criterion for the evaluation of methods, it is also conceivable, although not common, to define objective criteria for plausibility referring to literature or databases, and to evaluate the plausibility of the methods' results for several real datasets. Such an evaluation would not be without practical and conceptual difficulties, but would in principle address the flaw of the interpretations of real data examples discussed in this paper in two ways. Firstly, the evaluation would base on several datasets rather than on a single anecdotal dataset. Secondly, by defining objective criteria for plausibility one would reduce the RDF affecting interpretations.

Increasing awareness for the \lq\lq storytelling fallacy'' is especially important since the emergence of large language models (LLMs) in recent years. This stems from the fact that the flexibility researchers have when telling \lq\lq stories'' increases with the rise of user-friendly LLMs, which make it easier to generate literature-based consistent and plausible \lq\lq stories'' supporting the results of statistical methods---without even having to seek expert advice.

Based on the considerations outlined in this paper and previous literature on the design of methodological comparison studies, we formulate the following tentative recommendations. Real data applications are important and should remain an important part of methodological evaluations for illustrative purposes. However, anecdotal stories based on domain knowledge supporting the results of methods should not be considered as reliable evidence in favor of one or the other method---with only few exceptions involving several datasets and objective criteria. More generally, it is recommended to interpret real data applications in combination with those of simulation studies, and to abandon the \lq\lq one beats them all philosophy'' \cite{strobl2024against}. It should be acknowledged that no method is expected to yield uniformly \lq\lq better results''  in all situations. Relaxing the implicit expectation (of, e.g., editors and reviewers) that new methods should work clearly better than existing ones in all respects would certainly have a positive impact in terms of the incentive structure towards less biased interpretations.

\section*{Acknowledgments}

This study was partly funded by DFG grants BO3139/7-2 and BO3139/9-1 to Anne-Laure Boulesteix.

\section*{Author contributions}

Conceptualization: MM, ALB; Formal analysis: MM; Funding acquisition: ALB; Investigation: MM, FW, TW; Project Administration: ALB; Supervision: ALB; Writing – original draft: MM, ALB; Writing – review \& editing: MM, ALB, FW, TW.

\section*{Conflict of interest}

The authors have declared no conflict of interest. 

\section*{Data Availability Statement}

The data that support the findings of this study are not publicly available due to privacy or ethical restrictions.

\printbibliography

@article{boutron2018misrepresentation,
  author   = {Isabelle Boutron and Philippe Ravaud},
  title    = {Misrepresentation and distortion of research in biomedical literature},
  journal  = {Proceedings of the National Academy of Sciences},
  year     = {2018},
  volume   = {115},
  number   = {11},
  pages    = {2613-2619},
  url      = {https://www.pnas.org/doi/abs/10.1073/pnas.1710755115},
}

@Article{manousaki2020genome,
  author    = {Manousaki, Despoina and Mitchell, Ruth and Dudding, Tom and Haworth, Simon and Harroud, Adil and Forgetta, Vincenzo and Shah, Rupal L and Luan, Jian’an and Langenberg, Claudia and Timpson, Nicholas J and Richards, J. Brent},
  title     = {Genome-wide association study for vitamin D levels reveals 69 independent loci},
  journal   = {The American Journal of Human Genetics},
  year      = {2020},
  volume    = {106},
  number    = {3},
  pages     = {327--337},
  url = {https://doi.org/10.1016/j.ajhg.2020.01.017}
}

@article{strobl2024against,
author = {Strobl, Carolin and Leisch, Friedrich},
title = {Against the “one method fits all data sets” philosophy for comparison studies in methodological research},
journal = {Biometrical Journal},
volume = {66},
number = {1},
pages = {2200104},
doi = {https://doi.org/10.1002/bimj.202200104},
year = {2024}
}

@article{boulesteix2015publication,
author = {Anne-Laure Boulesteix and Veronika Stierle and Alexander Hapfelmeier},
title ={Publication Bias in Methodological Computational Research},

journal = {Cancer Informatics},
volume = {14s5},
year = {2015},
url = {https://doi.org/10.4137/CIN.S30747}
}

@article{ullmann2023over,
  title={Over-optimistic evaluation and reporting of novel cluster algorithms: An illustrative study},
  author={Ullmann, Theresa and Beer, Anna and H{\"u}nem{\"o}rder, Maximilian and Seidl, Thomas and Boulesteix, Anne-Laure},
  journal={Advances in Data Analysis and Classification},
  volume={17},
  number={1},
  pages={211--238},
  year={2023},
  url={https://doi.org/10.1007/s11634-022-00496-5}
}

@article{jelizarow2010over,
    author = {Jelizarow, Monika and Guillemot, Vincent and Tenenhaus, Arthur and Strimmer, Korbinian and Boulesteix, Anne-Laure},
    title = {Over-optimism in bioinformatics: an illustration},
    journal = {Bioinformatics},
    volume = {26},
    number = {16},
    pages = {1990--1998},
    year = {2010},
    url = {https://doi.org/10.1093/bioinformatics/btq323}
}

@article{pawel2024pitfalls,
author = {Pawel, Samuel and Kook, Lucas and Reeve, Kelly},
title = {Pitfalls and potentials in simulation studies: Questionable research practices in comparative simulation studies allow for spurious claims of superiority of any method},
journal = {Biometrical Journal},
volume = {66},
number = {1},
pages = {2200091},
url = {https://onlinelibrary.wiley.com/doi/abs/10.1002/bimj.202200091},
year = {2024}
}

@article{dror2020cognitive,
author = {Dror, Itiel E.},
title = {Cognitive and Human Factors in Expert Decision Making: Six Fallacies and the Eight Sources of Bias},
journal = {Analytical Chemistry},
volume = {92},
number = {12},
pages = {7998--8004},
year = {2020},
url = {https://doi.org/10.1021/acs.analchem.0c00704}
}

@article{o2019expert,
author = {Anthony O’Hagan},
title = {Expert Knowledge Elicitation: Subjective but Scientific},
journal = {The American Statistician},
volume = {73},
number = {sup1},
pages = {69--81},
year = {2019},
url = {https://doi.org/10.1080/00031305.2018.1518265}
}

@misc{brownstein2018perspective,
      title={Perspective from the Literature on the Role of Expert Judgment in Scientific and Statistical Research and Practice}, 
      author={Naomi C Brownstein},
      year={2018},
      eprint={1809.04721},
      archivePrefix={arXiv},
      primaryClass={stat.OT},
      url={https://arxiv.org/abs/1809.04721}, 
}

@article {kaptchuk2003effect,
	author = {Kaptchuk, Ted J},
	title = {Effect of interpretive bias on research evidence},
	volume = {326},
	number = {7404},
	pages = {1453--1455},
	year = {2003},
	publisher = {BMJ Publishing Group Ltd},
	URL = {https://www.bmj.com/content/326/7404/1453},
	journal = {BMJ}
}

@article{baker20161,
  title={1,500 scientists lift the lid on reproducibility},
  author={Baker, Monya},
  journal={Nature},
  volume={533},
  pages = {452--454},  
  year={2016},
  url = {https://doi.org/10.1038/533452a}
}

@article{gelman2014statistical,
  title={The statistical crisis in science},
  author={Gelman, Andrew and Loken, Eric},
  journal={American Scientist},
  volume={102},
  number={6},
  pages={460--465},
  year={2014}
}

@Article{loken2017measurement,
  author   = {Eric Loken and Andrew Gelman},
  title    = {Measurement error and the replication crisis},
  journal  = {Science},
  year     = {2017},
  volume   = {355},
  number   = {6325},
  pages    = {584--585},
  url      = {https://www.science.org/doi/abs/10.1126/science.aal3618},
}

@article{simmons2011false,
  title={False-positive psychology: Undisclosed flexibility in data collection and analysis allows presenting anything as significant},
  author={Simmons, Joseph P and Nelson, Leif D and Simonsohn, Uri},
  journal={Psychological Science},
  volume={22},
  number={11},
  pages={1359--1366},
  year={2011},
  publisher={Sage Publications Sage CA: Los Angeles, CA},
  doi={10.1177/0956797611417632}
}

@article{nuzzo2015fooling,
  title={Fooling ourselves},
  author={Nuzzo, Regina},
  journal={Nature},
  volume={526},
  pages={182--185},
  year={2015},
  url={https://doi.org/10.1038/526182a}
}

@article{hoffmann2021multiplicity,
  title={The multiplicity of analysis strategies jeopardizes replicability: lessons learned across disciplines},
  author={Hoffmann, Sabine and Sch{\"o}nbrodt, Felix and Elsas, Ralf and Wilson, Rory and Strasser, Ulrich and Boulesteix, Anne-Laure},
  journal={Royal Society Open Science},
  volume={8},
  number={4},
  pages={201925},
  year={2021},
  url={https://doi.org/10.1098/rsos.201925}
}

@article{ioannidis2005most,
  title={Why most published research findings are false},
  author={Ioannidis, John PA},
  journal={PLoS Medicine},
  volume={2},
  number={8},
  pages={e124},
  year={2005},
  url={https://doi.org/10.1371/journal.pmed.0020124}
}

@incollection{oswald2004confirmation,
  author    = {Oswald, Margit E. and Grosjean, Sabine},
  title     = {Confirmation Bias},
  booktitle = {Cognitive Illusions: A Handbook on Fallacies and Biases in Thinking, Judgement and Memory},
  editor    = {Pohl, Rüdiger F.},
  publisher = {Psychology Press},
  address   = {Hove and New York},
  year      = {2004},
  pages={79--96}
}

@article{bowden2015egger,
    author = {Bowden, Jack and Davey Smith, George and Burgess, Stephen},
    title = {Mendelian randomization with invalid instruments: effect estimation and bias detection through Egger regression},
    journal = {International Journal of Epidemiology},
    volume = {44},
    number = {2},
    pages = {512-525},
    year = {2015},
    url = {https://doi.org/10.1093/ije/dyv080},
}

@article{sanderson2019examination,
    author = {Sanderson, Eleanor and Davey Smith, George and Windmeijer, Frank and Bowden, Jack},
    title = {An examination of multivariable Mendelian randomization in the single-sample and two-sample summary data settings},
    journal = {International Journal of Epidemiology},
    volume = {48},
    number = {3},
    pages = {713--727},
    year = {2019},
    url = {https://doi.org/10.1093/ije/dyy262}
}

@article{verbanck2018detection,
  title={Detection of widespread horizontal pleiotropy in causal relationships inferred from Mendelian randomization between complex traits and diseases},
  author={Verbanck, Marie and Chen, Chia-Yen and Neale, Benjamin and Do, Ron},
  journal={Nature Genetics},
  volume={50},
  number={5},
  pages={693-–698},
  year={2018},
  url={ https://doi.org/10.1038/s41588-018-0099-7}
}

@article{woehrle2024point,
author = {Woehrle, Tobias and Pfeiffer, Florian and Mandl, Maximilian M. and Sobtzick, Wolfgang and Heitzer, Jörg and Krstova, Alisa and Kamm, Luzie and Feuerecker, Matthias and Moser, Dominique and Klein, Matthias and Aulinger, Benedikt and Dolch, Michael and Boulesteix, Anne-Laure and Lanz, Daniel and Choukér, Alexander},
title = {Point-of-care breath sample analysis by semiconductor-based E-Nose technology discriminates non-infected subjects from SARS-CoV-2 pneumonia patients: a multi-analyst experiment},
journal = {MedComm},
volume = {5},
number = {11},
pages = {e726},
url = {https://onlinelibrary.wiley.com/doi/abs/10.1002/mco2.726},
year = {2024}
}

@article{kramer2009regularized,
  title={Regularized estimation of large-scale gene association networks using graphical Gaussian models},
  author={Kr{\"a}mer, Nicole and Sch{\"a}fer, Juliane and Boulesteix, Anne-Laure},
  journal={BMC Bioinformatics},
  volume={10},
  number={384},
  year={2009},
  url={https://doi.org/10.1186/1471-2105-10-384}
}

@article{jiang2024gasotransmitter,
author = {Jiang, Simin and Chen, Haijie and Shen, Pu and Zhou, Yumou and Li, Qiaoyu and Zhang, Jing and Chen, Yahong},
title = {Gasotransmitter Research Advances in Respiratory Diseases},
journal = {Antioxidants \& Redox Signaling},
volume = {40},
number = {1-3},
pages = {168--185},
year = {2024},
url = {https://doi.org/10.1089/ars.2023.0410}
}

@article{citi2020anti,
author = {Citi, Valentina and Martelli, Alma and Brancaleone, Vincenzo and Brogi, Simone and Gojon, Gabriel and Montanaro, Rosangela and Morales, Guillermo and Testai, Lara and Calderone, Vincenzo},
title = {Anti-inflammatory and antiviral roles of hydrogen sulfide: Rationale for considering H2S donors in COVID-19 therapy},
journal = {British Journal of Pharmacology},
volume = {177},
number = {21},
pages = {4931--4941},
url = {https://bpspubs.onlinelibrary.wiley.com/doi/abs/10.1111/bph.15230},
year = {2020}
}

@article{heinze2024phases,
author = {Heinze, Georg and Boulesteix, Anne-Laure and Kammer, Michael and Morris, Tim P. and White, Ian R. and the Simulation Panel of the STRATOS initiative},
title = {Phases of methodological research in biostatistics—Building the evidence base for new methods},
journal = {Biometrical Journal},
volume = {66},
number = {1},
pages = {2200222},
url = {https://onlinelibrary.wiley.com/doi/abs/10.1002/bimj.202200222},
year = {2024}
}

@article{mokry2015vitamin,
  title={Vitamin D and risk of multiple sclerosis: a Mendelian randomization study},
  author={Mokry, Lauren E and Ross, Stephanie and Ahmad, Omar S and Forgetta, Vincenzo and Smith, George Davey and Leong, Aaron and Greenwood, Celia MT and Thanassoulis, George and Richards, J Brent},
  journal={PLoS Medicine},
  volume={12},
  number={8},
  pages={e1001866},
  year={2015},
  url={https://doi.org/10.1371/journal.pmed.1001866}
}

@article{burgess2013mendelian,
author = {Burgess, Stephen and Butterworth, Adam and Thompson, Simon G.},
title = {Mendelian Randomization Analysis With Multiple Genetic Variants Using Summarized Data},
journal = {Genetic Epidemiology},
volume = {37},
number = {7},
pages = {658--665},
url = {https://onlinelibrary.wiley.com/doi/abs/10.1002/gepi.21758},
year = {2013}
}

@article{niessl2024explaining,
author = {Nießl, Christina and Hoffmann, Sabine and Ullmann, Theresa and Boulesteix, Anne-Laure},
title = {Explaining the optimistic performance evaluation of newly proposed methods: A cross-design validation experiment},
journal = {Biometrical Journal},
volume = {66},
number = {1},
pages = {2200238},
url = {https://onlinelibrary.wiley.com/doi/abs/10.1002/bimj.202200238},
year = {2024}
}

@article{niessl2022over,
author = {Nießl, Christina and Herrmann, Moritz and Wiedemann, Chiara and Casalicchio, Giuseppe and Boulesteix, Anne-Laure},
title = {Over-optimism in benchmark studies and the multiplicity of design and analysis options when interpreting their results},
journal = {WIREs Data Mining and Knowledge Discovery},
volume = {12},
number = {2},
pages = {e1441},
url = {https://wires.onlinelibrary.wiley.com/doi/abs/10.1002/widm.1441},
year = {2022}
}

@article{boulesteix2013plea,
  title={A plea for neutral comparison studies in computational sciences},
  author={Boulesteix, Anne-Laure and Lauer, Sabine and Eugster, Manuel JA},
  journal={PLoS ONE},
  volume={8},
  number={4},
  pages = {e61562},
  year={2013},
  url={https://doi.org/10.1371/journal.pone.0061562}
}

@article{buchka2021optimistic,
  title={On the optimistic performance evaluation of newly introduced bioinformatic methods},
  author={Buchka, Stefan and Hapfelmeier, Alexander and Gardner, Paul P and Wilson, Rory and Boulesteix, Anne-Laure},
  journal={Genome Biology},
  volume={22},
  number={152},
  year={2021},
  url={https://doi.org/10.1186/s13059-021-02365-4}
}

@article{mandl2025outlierdetectionmendelianrandomisation,
      title={Outlier Detection in Mendelian Randomisation}, 
      author={Mandl, Maximilian M. and Boulesteix, Anne-Laure and Burgess, Stephen and Zuber, Verena},
  journal={Statistics in Medicine (conditionally accepted)},
  year={2025},
      url={https://arxiv.org/abs/2502.14716}, 
}

@article{gelman2017beyond, 
   author = {Gelman, Andrew and Hennig, Christian}, 
   title = {Beyond Subjective and Objective in Statistics}, 
   journal = {Journal of the Royal Statistical Society Series A: Statistics in Society}, 
   volume = {180}, 
   number = {4}, 
   pages = {967--1033}, 
   year = {2017}, 
   doi = {10.1111/rssa.12276}, 
   url = {https://doi.org/10.1111/rssa.12276} 
}

@article{lorincz2022,
title = {The role of cholesterol metabolism in multiple sclerosis: From molecular pathophysiology to radiological and clinical disease activity},
journal = {Autoimmunity Reviews},
volume = {21},
number = {6},
pages = {103088},
year = {2022},
issn = {1568-9972},
doi = {https://doi.org/10.1016/j.autrev.2022.103088},
url = {https://www.sciencedirect.com/science/article/pii/S1568997222000581},
author = {Balazs Lorincz and Elizabeth C. Jury and Michal Vrablik and Murali Ramanathan and Tomas Uher},
}

@article{murali2022,
author = {Murali, N. and Browne, R. W. and Fellows Maxwell, K. and Bodziak, M. L. and Jakimovski, D. and Hagemeier, J. and Bergsland, N. and Weinstock-Guttman, B. and Zivadinov, R. and Ramanathan, M.},
title = {Cholesterol and neurodegeneration: longitudinal changes in serum cholesterol biomarkers are associated with new lesions and gray matter atrophy in multiple sclerosis over 5 years of follow-up},
journal = {European Journal of Neurology},
volume = {27},
number = {1},
pages = {188-e4},
url = {https://onlinelibrary.wiley.com/doi/abs/10.1111/ene.14055},
year = {2020}
}

@book{hastie01statisticallearning,
  address = {New York, NY, USA},
  author = {Hastie, Trevor and Tibshirani, Robert and Friedman, Jerome},
  publisher = {Springer New York Inc.},
  series = {Springer Series in Statistics},
  title = {The Elements of Statistical Learning},
  year = 2001
}

\end{document}